\documentclass[%
 aip,
cp,  
 amsmath,amssymb,
 reprint,%
]{revtex4-2}
\usepackage{fancyhdr}
\usepackage{graphicx}
\usepackage{dcolumn}
\usepackage{bm}

\usepackage[utf8]{inputenc}
\usepackage[T1]{fontenc}

\usepackage{mathptmx} 

\usepackage{lineno}
\begin{document}
\pdfoutput=1
\title{A Fractional Viscoelastic Model of the Axon in Brain White Matter}

\author{Parameshwaran Pasupathy} 
 \email{parameshwaran.pasupathy@rutgers.edu}

\affiliation{
 Department of Mechanical and Aerospace Engineering, Rutgers \, The State University of New Jersey, 98 Brett Road, Piscataway \, NJ 08854 \ \\
}

\author{John G Georgiadis}%
\email{jgeorgia@iit.edu}
\affiliation{ Department of Biomedical Engineering, Illinois Institute of Technology, 3255 S. Dearborn St., Wishnick Hall 314, Chicago \, IL 60616 \\}

\author{Assimina A Pelegri}
\email[Corresponding author: ]{pelegri@rutgers.edu}
\affiliation{
	Department of Mechanical and Aerospace Engineering, Rutgers \, The State University of New Jersey, 98 Brett Road, Piscataway \, NJ 08854 \ \\
}


\begin{abstract}
Traumatic axonal injury occurs when loads experienced on the tissue-scale are transferred to the individual axons. Mechanical characterization of axon deformation especially under dynamic loads however is extremely difficult owing to their viscoelastic properties. The viscoelastic characterization of axon properties that are based on interpretation of results from in vivo brain Magnetic Resonance Elastography (MRE) are dependent on the specific frequencies used to generate shear waves with which measurements are made. In this study, we aim to develop a fractional viscoelastic model to characterize the time dependent behavior of the properties of the axons in a composite white matter (WM) model. The viscoelastic powerlaw behavior observed at the tissue level is assumed to exist across scales, from the continuum macroscopic level to that of the microstructural realm of the axons. The material parameters of the axons and glia are fitted to a springpot model. The 3D fractional viscoelastic springpot model is implemented within a finite element framework. The constitutive equations defining the fractional model are coded using a vectorized user defined material (VUMAT) subroutine in ABAQUS finite element software. Using this material characterization,  Representative Volume Elements (RVE) of axons embedded in glia with periodic boundary conditions are developed and subjected to a creep displacement boundary condition. The homogenized orthotropic fractional material properties of the axon-matrix system as a function of the volume fraction of axons in the ECM are extracted by solving the inverse problem.
\end{abstract}

\maketitle
\fancypagestyle{alim}{\fancyhf{}\renewcommand{\headrulewidth}{0pt}\fancyhead[L]{Accepted for publication in IC-MSQUARE, 11th Int'l Conference on Mathematical Modeling in Physical Sciences.\\
AIP Conference Proceedings Volume 2872}}
\section{\label{sec:level1}Introduction}

As in a myriad other areas, the finite element method (FEM) has shown much promise in characterizing brain tissue response to mechanical loading and in understanding injury biomechanics. The method allows for characterization of brain tissue along multiple length scales; from whole head models that can be used to design safer helmets [\onlinecite{AARE2007596}] to micromechanical models at the level of the axons and microtubules [\onlinecite{pan2013finite}]. The major challenge, however, in effectively predicting the mechanics of brain tissue lies in capturing its mechanical properties accurately across multiple spatial and temporal scales. This is extremely difficult as the mechanical response of brain tissue is highly anisotropic and nonlinear[\onlinecite{abolfathi2009}]. Moreover, the brain as an organ is completely enclosed and is hard to probe. These mechanical properties play a vital role during trauma related events, and can be used to predict the onset of neurodegenerative diseases. \\

\thispagestyle{alim} 
\noindent Constituting approximately 50 percent of the brain, white matter is a significant region in disease onset and senescence [\onlinecite{sullivan2021}]. Axonal damage in the corpus callosum of the white matter has been identified as the leading cause of traumatic brain injury (TBI), with excessive tensile strain postulated as the underlying mechanism [\onlinecite{arbogast1998material}]. Considering the fact that brain tissue is extremely soft, viscous and its mechanical response: non-linear, a number of researchers have used hyperelastic as well as viscoelastic material models to characterize its mechanics. Meaney [\onlinecite{meaney2003}] proposed an analytical structural model to formulate structure-property relationships for high-directional tortuous axons using different hyperelastic strain energy functions. Arbogast et al. [\onlinecite{arbogast1998material}] performed experiments on guinea optic fiber nerves and developed a viscoelastic fiber reinforced composite model of the axons in an extracellular matrix (ECM) in the frequency domain. Montanino et al. [\onlinecite{montanino2018}] developed a microstructural model of the axon and its substructures under different strain rates to study axonal injury mechanisms. Javed et al. [\onlinecite{JAVID2014290}] used the genetic algorithm optimization procedure to determine the homogenized prony series parameters for representative volume elements (RVE) of axons in the ECM. The research was based on relaxation tests performed on the porcine brain white matter. 
Pan et al. [\onlinecite{pan2011transition}] developed a transitional micromechanical model that captures the transition of axons from non-affine-dominated kinematics at low stretch levels to affine kinematics at high stretch levels using the Ogden hyperelastic material model. Sullivan et al. [\onlinecite{sullivan2021}] developed a triphasic unidirectional composite model consisting of axons, myelin and ECM and derived homogenized viscoelastic material properties under steady state dynamics. \\

\noindent An increasing number of research studies indicate that the viscoelastic response of brain white matter can be phenomenologically explained by a power-law behavior [\onlinecite{sack2009impact, sack2013structure, nicolas2018biomechanical, kurt2019optimization}]. Sack et al. [\onlinecite{sack2009impact}] used multifrequency magnetic resonance elastography (MRE) measurements of the human brain in 55 volunteers and developed a power law model that predicted the effect of aging on the human brain. Kurt et al. [\onlinecite{kurt2019optimization}] developed a protocol to calculate optimal frequency sets and determine power-law parameters separately for the entire brain, white matter and grey matter. Nicolas et al. [\onlinecite{nicolas2018biomechanical}] performed ex vivo brain experiments using ultrasound shear wave spectroscopy and determined its mechanical parameters by fitting a power-law model. In this study, we posit that the viscoelastic power-law behavior observed at the tissue level exists across all length scales, from the continuum macroscopic level to that of the microstructural realm of the axons. A viscoelastic power-law model of a springpot is applied to the axons and the ECM. The material parameters for the springpot are obtained via a logistic regression analysis. A 3D fractional viscoelastic springpot model is implemented within a finite element framework. Homogenized fractional viscoelastic parameters as a function of the volume fractions of axons embedded in the ECM are then derived.  

\section{\label{sec:level1}Materials and Method}

The stress strain response for a viscoelastic material at time \(t\) with the action of stress beginning at time \(\tau\) can be described using a convolution integral of the form

\begin{equation}
	\sigma(t) = \int_{0}^{t} G(t- \tau) \frac{d \epsilon(\tau)}{d \tau} \hspace{2mm}d\tau,
\end{equation}

\noindent and the strain as a function of stress is determined by 

\begin{equation}
		\epsilon(t) = \int_{0}^{t} 	J(t- \tau) \frac{d \sigma(\tau)}{d \tau} \hspace{2mm}d\tau.
\end{equation}

\noindent where $G(t-\tau)$ and $J(t-\tau)$ are the relaxation and creep modulus of the material, respectively. The relaxation modulus of a viscoelastic material with constants $C_{\beta}$ and \(\beta\), described by a power-law is,

\begin{equation}
	G(t) = C_{\beta} t^{-\beta}
\end{equation}

\noindent The parameters $C_{\beta}$ and \(\beta\) for the axons and ECM are determined by logistic regression analysis using frequency dependent data from porcine optic nerve fiber experiments published in [\onlinecite{arbogast1998material}]. The power-law model in the frequency domain can be written as, 

\begin{equation}
	G(\omega) = \kappa (i \omega)^\beta = G' + iG" = \Re(G(\omega)) + i\Im(G(\omega)),
\end{equation}

\noindent where $G(\omega)$ is the complex relaxation modulus in the frequency domain. \(\kappa\) and \(\beta\) are constants. $G'$ is the real part of the complex modulus and is known as the storage modulus.  $G"$ is the imaginary part of the complex modulus and is known as the loss modulus (see Figure 1). A cost function is defined for the axons and ECM as, 

\begin{equation}
		E = \frac{1}{2m} \sum_m ((\ln(\kappa) + \beta \ln(\omega) - \Re(G^*))^2 + (\beta \frac{\pi}{2} - \Im(G^*))^2 ),
\end{equation}

\noindent where $G^*$ is the complex shear modulus from [\onlinecite{arbogast1998material}]. m is the number of input frequency points. Parameters obtained by minimizing the cost function is shown in Table 1. Table 2 illustrates the model's power-law parameters in comparison with published data [\onlinecite{sack2009impact},  \onlinecite{kurt2019optimization}]. \\

\begin{figure}[h]
	\includegraphics[scale=1.2]{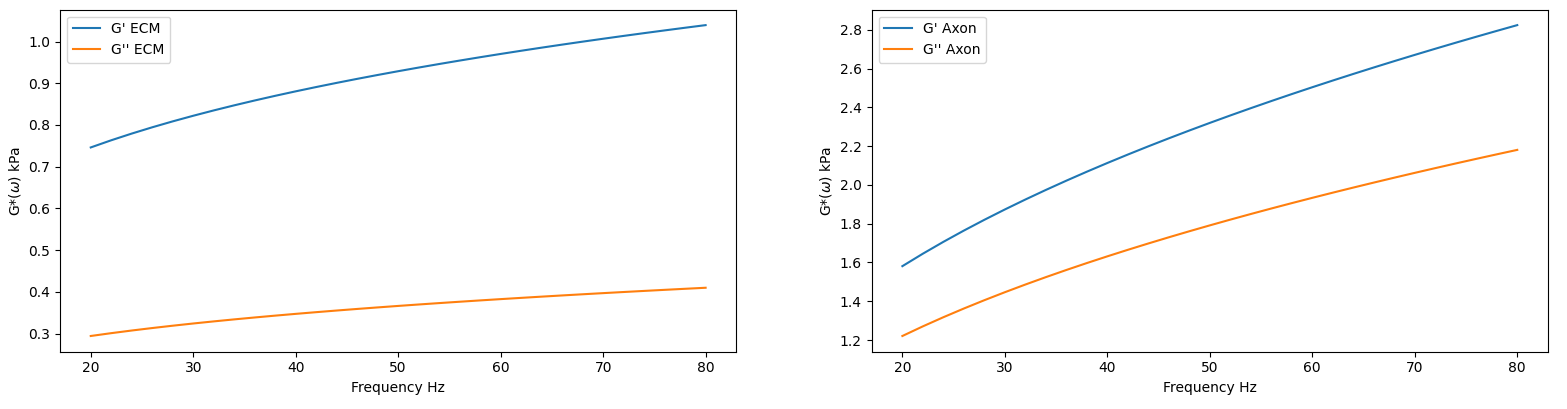}
	\caption{\label{fig:G*} A plot of the storage and loss modulus for the axons and the ECM as a function of the frequency using power-law parameters in Table 1.}
\end{figure}
\begin{table}
	\caption{\label{tab:table1}Power-law parameters for axons and ECM using logistic regression analysis on [\onlinecite{sack2009impact}]}
	\begin{ruledtabular}
		\begin{tabular}{lcr}
			Component &$\kappa (kPa \cdot s^\beta)$&$\beta$\\
			\hline
			Axon & 0.2641 & 0.419\\
			ECM & 0.2525 & 0.239\\
		\end{tabular}
	\end{ruledtabular}
\end{table}

\begin{table}[ht!]
	\caption{A comparison of the power-law exponent $\beta$ with published data}
	\begin{center}
		\begin{tabular}{lc}
			\hline \hline
			Author     & \begin{tabular}[c]{@{}c@{}}Powerlaw Exponent\\ \(\beta\)\end{tabular}       \\ 
			\hline
			BWM\footnotemark[1] & 0.264                                                      \\ \hline
			BWM\footnotemark[2] & 0.339 \\ \hline
			Axon\footnotemark[3] & 0.419 \\ \hline
			ECM\footnotemark[3] & 0.239 \\ \hline \hline
		\end{tabular}
	\end{center}
\footnotetext[1]{Sack et al. [\onlinecite{sack2009impact}]}
\footnotetext[1]{Kurt et al. [\onlinecite{kurt2019optimization}]}
\footnotetext[3]{Current study}
\end{table}

\begin{figure}[ht!]
	\includegraphics[scale=1.2]{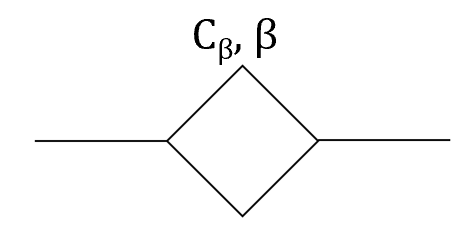}
	\caption{\label{fig:sp} A schematic representation of a springpot.}
\end{figure}

\noindent The mechanical response of a material with a relaxation modulus described in equation (3) can be represented by a Scott-Blair linear viscoelastic model commonly known as a springpot [\onlinecite{bonfanti2020fractional}]. The model's physical behavior is an intermediate to that of a spring and a viscous dashpot, see Figure 2. The mathematical implementation of the springpot is obtained using the notion of derivatives of non-integer order or fractional derivatives. The mathematical relationship can thus be written as, 

\begin{equation}
	\sigma(t) =  C_\beta \frac{d^\beta \epsilon(t)}{dt^\beta} \hspace{1mm}	\forall \hspace{1mm} (0 \leq \beta \leq 1). 
\end{equation}

\noindent A generalization of derivatives of non-integer orders can be obtained in a branch of mathematics called fractional calculus using the Caputo derivative, which is given by 

\begin{equation}
	\frac{d^\beta \epsilon(t)}{dt^\beta} = \frac{1}{\Gamma(1 - \beta)}\int_{0}^{t}  (t-\tau)^{-\beta}\frac{d \epsilon(\tau)}{d \tau} \hspace{2mm}d\tau,
\end{equation}

\noindent where $\Gamma(\cdot)$ is the gamma function. Substituting the power-law equation (3) into (6), (7) and applying a Fourier transformation yields \\

\begin{equation}
	\sigma(\omega) = C_\beta (i \omega)^\beta \epsilon(\omega)
\end{equation}

\noindent Comparing the relaxation modulus in (4) and (8), we see that $\kappa = C_\beta$. Thus, the stress-strain relationship for a springpot can be written as,

\begin{equation}
	\sigma(t) = \frac{\kappa}{\Gamma(1 - \beta)}\int_{0}^{t}  (t-\tau)^{-\beta}\frac{d \epsilon(\tau)}{d \tau} \hspace{2mm}d\tau
\end{equation}

\noindent The fractional springpot model for the microstructure of CNS white matter is implemented using the finite element method in ABAQUS finite element solver. For a 3D finite element model, the relaxation matrix can be split into its volumetric and deviatoric components [\onlinecite{alotta2017behavior}],

\begin{equation}
	G_{ijkm}(t) = (K_R(t) - \frac{2}{3} G_{R}(t)) \delta_{ij} \delta_{km} + G_{R}(t)(\delta_{ik} \delta_{jm} + \delta_{im} \delta_{jk}),
\end{equation}

\noindent where $\delta_{ij}$ is the Kronecker delta function. $G_R(t)$ and $K_R(t)$ are the deviatoric and volumetric power law functions respectively. Substituting (10) into (9) and rewriting terms of the power-law coefficient $\kappa$, in terms the volumetric and deviatoric components, $K_\beta$ and $G_\beta$, the stress-strain equation becomes,

\begin{equation}
	\sigma_{ij} = \frac{1}{\Gamma(1-\beta)}\int_{0}^{t} (K_\beta - \frac{2}{3} G_\beta) \delta_{ij} \dot{\epsilon_{kk}}(t) dt + \frac{1}{\Gamma(1-\beta)}\int_{0}^{t} G_\beta (\dot{\epsilon_{ij}}(t) + \dot{\epsilon_{ji}}(t)) dt
\end{equation}

\noindent Alotta et al. [\onlinecite{alotta2018finite}] developed a 3D fractional viscoelastic user material (UMAT) subroutine in Abaqus. The fractional viscoelastic model is implemented numerically using the Grunwald- Letnikov Operator [\onlinecite{199941}, \onlinecite{1999199}]. For a 3D state of stress, $\sigma\left(t\right)={[\sigma}_{11}\ \sigma_{22\ }\sigma_{33}\ \tau_{23}\ \tau_{31}\ \tau_{12}]$, the stress in any direction is given by,

\begin{equation}
\sigma_{ij}^{k+1}={(K}_\beta\ -\ \frac{2G_\beta}{3})\left(\frac{1}{\Delta t}\right)^\beta\sum_{l=1}^{k+1}\varphi_l\epsilon_{kk}\left(\left(k+1-l\right)\Delta t\right)+2G_\beta\left(\frac{1}{\Delta t}\right)^\beta\sum_{l=1}^{k+1}\varphi_l\epsilon_{ij}((k+1\ -\ l)\Delta t),
\end{equation}

\noindent where $k = \frac{Totaltime}{\Delta t}$, is the number of iterations, $\epsilon_{kk}$ is the volumetric strain and $\phi_l$ are the Grunwald coefficients which can be calculated as [\onlinecite{1999199}],

\begin{equation}
		\varphi_{l+1} = \frac{(k - 1 -\beta)}{k} \varphi_l,	 \hspace{5mm} \varphi_1 = 1.
\end{equation}

\noindent Note that the calculation of stress at any given increment requires storing and accessing the history of strains for all previous increments. This makes the simulation of large models in an explicit integration scheme computationally expensive.\\

\begin{figure}[ht!]
	\includegraphics[scale=0.7]{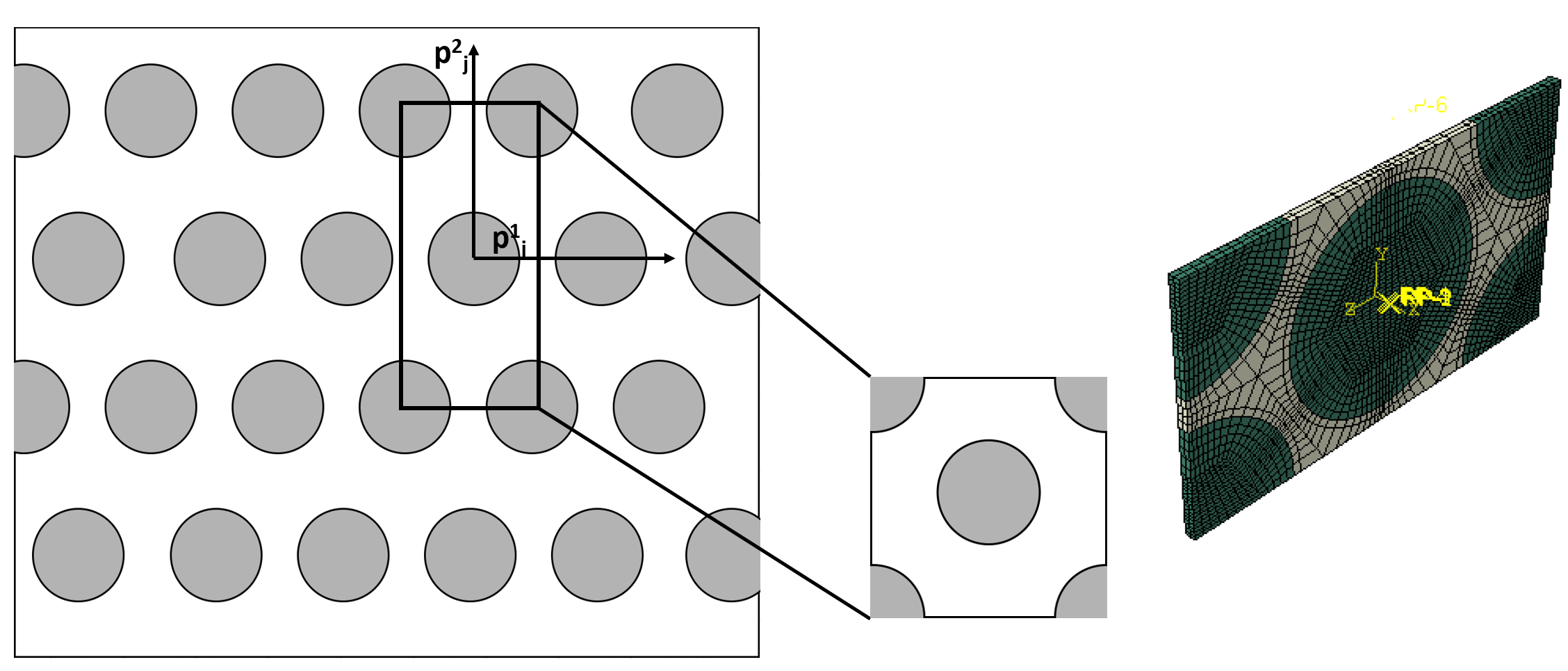}
	\caption{\label{fig:RVE} \textsl{Left} A schematic representation of a periodic geometry of axons in the ECM. \textsl{Right} A finite element model of the hexagonally packed RVE of two axons embedded in the ECM.}
\end{figure}

\noindent Hexagonally packed RVE models of the axons embedded in the ECM of varying volume fractions is developed with a periodic mesh as shown in Figure 3. The diameter of the axon in the RVE is 10 $\mu$m. Periodic boundary conditions on the nodes are imposed through equation constraints in ABAQUS [\onlinecite{tian2019periodic}, \onlinecite{omairey2019development}]. The far-field gradient is applied through the degrees of freedom of reference points (RP) [\onlinecite{omairey2019development}]. RP nodes are not attached to any of the elements in the model. Displacement controlled boundary conditions are applied using the following equation [\onlinecite{abq}], 

\begin{equation}
	u_i(x_j + p_j^\alpha) = u_i(x_j) + \frac{\partial u_i}{\partial x_j} p_j^\alpha,
\end{equation}

\noindent where $x_j$ is the coordinate,  $ p_j^\alpha$ is the $\alpha$th vector of periodicity, and $\frac{\partial u_i}{\partial x_j}$ is the far-field gradient of the displacement. The linear constraint equations in ABAQUS are of the form [\onlinecite{abq}], 

\begin{equation}
	A_1 u^p_i + A_2 u^q_j + \dots + A_n u^r_k = 0, 
\end{equation}

\noindent where $N$ is the number of terms in the equation, $u^P_i$ corresponds to displacement variable of node $P$ and degree of freedom $i$, and $A_n$ are the coefficients. \\

\noindent The finite element model of the RVE with the fractional viscoelastic VUMAT is subjected to a relaxation strain test in six directions. An explicit time integration technique is used to solve the FE model. The homogenized stresses extracted from the simulation of the RVE for each loading direction is fed to an optimization workflow. 

\begin{figure}[h]
	\includegraphics[scale=0.7]{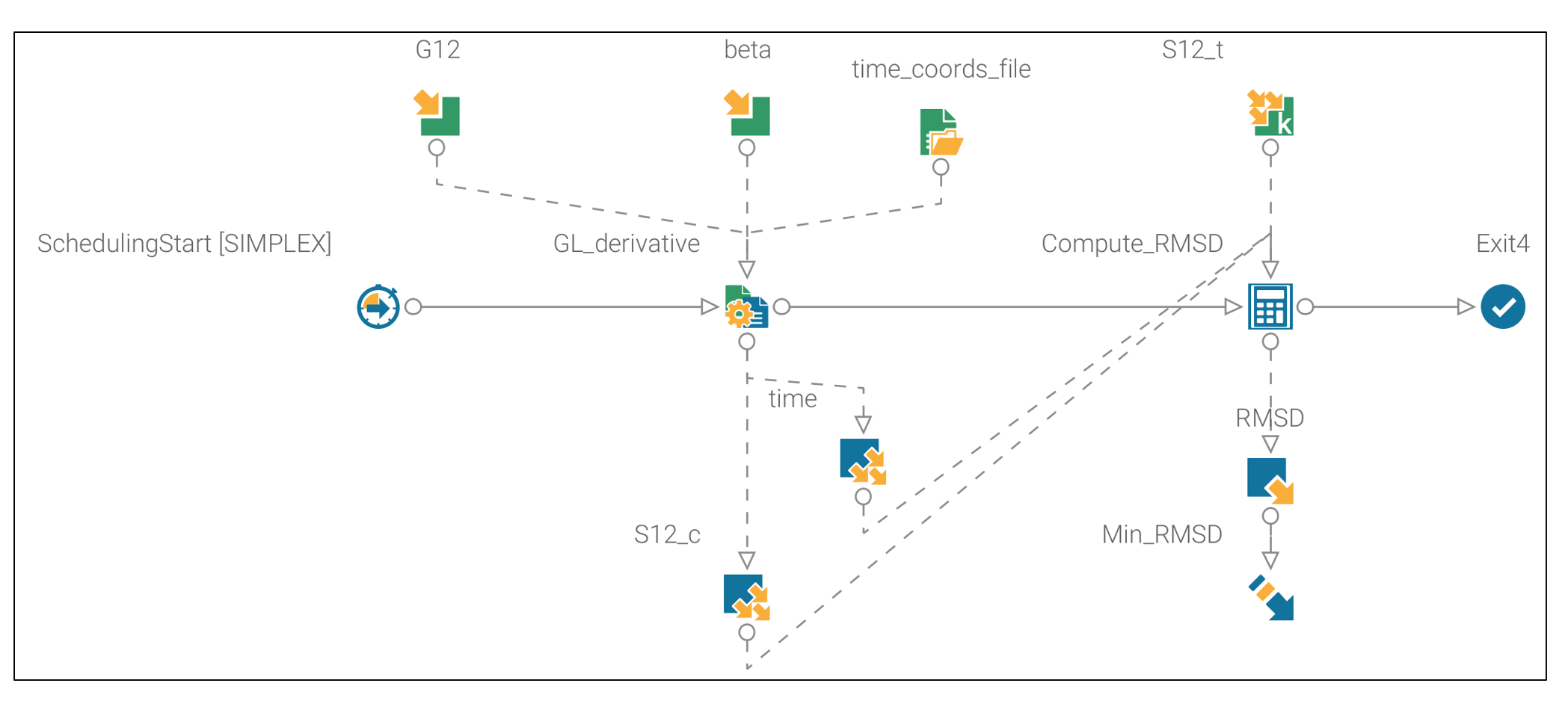}
	\caption{\label{fig:mf} Optimization workflow to extract optimal fractional properties for the RVE using simplex optimization implemented in modeFRONTIER}
\end{figure}

\noindent The workflow is implemented in modeFRONTIER optimization software, see Figure 4. A Nelder-Mead downhill simplex algorithm is used to determine the optimal parameters [\onlinecite{mf}]. Simplex is a geometric figure with $n+1$ vertices in an n-dimensional space. It compares the values of the objective function at $N+1$ vertices and gradually moves the polyhedron towards the optimal point by iteratively replacing the worst vertex with a point moved through the centroid of the remaining N points. The workflow computes the fractional viscoelastic stress for any input parameter and minimizes the root mean square (RMSD) between the computed stress and the homogenized stress from the FEM simulation. The optimization problem is formulated as shown in equation (16) where $J$ is the cost function, $\sigma_{sim}$ is the FEM solution and $\sigma_{c}$ is the computed stress. 

\begin{subequations}
	\begin{alignat}{2}
		&\!\min        &\qquad& J = \sqrt{\frac{1}{m}\sum_{n=1}^m \frac{(\sigma_{sim} - \sigma_{c})^2}{\sigma_{sim}^2}} \label{eq:optProb}\\
		 &\text{subject to} &      & 0 \leq \beta \leq 1 ; \hspace{3mm}m > 0 \label{eq:constraint1}
	\end{alignat}
\end{subequations}

\section{\label{sec:level1}Results and Discussion}

\noindent The implementation of the VUMAT algorithm is first verified using a single element test case under uniaxial tension. The single element model is subjected to a relaxation test with a maximum strain of 0.01, see Figures 5-6. The power-law related shear and bulk modulus parameters are obtained from [\onlinecite{alotta2018finite}, \onlinecite{hesammokri2019implementation}] (Table 3). The model is solved using an explicit time integration scheme. The analytical solution is obtained by substituting the relaxation strain into (11). The comparison of the analytical solution with the VUMAT solution is shown in figure 7. The VUMAT algorithm exactly reproduces the analytical solution.

\begin{table}[ht!]
	\caption{A comparison of the power-law exponent $\beta$ with published data}
	\begin{center}
		\begin{tabular}{lc}
			\hline \hline
			Parameter     & \begin{tabular}[c]{@{}c@{}}Value $(MPa \cdot s^\beta)$\\ \end{tabular}       \\ 
			\hline
			$K_\beta$ & 500.0 \\ \hline
			$G_\beta$ & 375.0 \\ \hline \hline
		\end{tabular}
	\end{center}
\end{table}

\noindent A drawback of using the fractional viscoelastic model is the need to store and retrieve the entire history of strains in order to calculate the stress at the end of each increment. Podlubny [\onlinecite{1999199}] describes a method known as the short memory principle which can be used to truncate the memory of strains required to compute the stress in equation (17). This reduces the accuracy of the solution and there are no set parameters to calculate the optimal memory length. The optimal memory length has to be determined for each load profile separately for the desired accuracy. Podlubny [\onlinecite{1999199}] illustrates a number of examples with an error estimate for truncating memory. Alota et al. [\onlinecite{alotta2018finite}] also discuss examples of truncated memory for explicit simulations. The reduced memory solution for the single element test case with different memory lengths is shown in Figure 8. For the relaxation test, all values of the memory length, L, reproduce the maximum stress in the model, see Figure 9. Deviation from the exact solution occurs in the relaxation phase of the stress for larger values of L. 

\begin{subequations}
	\begin{alignat}{2}
		&   &\qquad& \frac{d^\beta f}{dt^\beta} = \left(\frac{1}{\Delta t}\right)^\beta\sum_{l=1}^{M+1}\varphi_lf\left(\left(k+1-l\right)\Delta t\right)\label{eq:optProb}\\
		&\text{where} &      & M = \text{min} \{k, \frac{L}{\Delta t}\}, \hspace{2mm} L = \textsl{Memory Length}
	\end{alignat}
\end{subequations}

\begin{figure}[ht]
	\begin{minipage}{0.35\linewidth}
		\includegraphics[width=\textwidth]{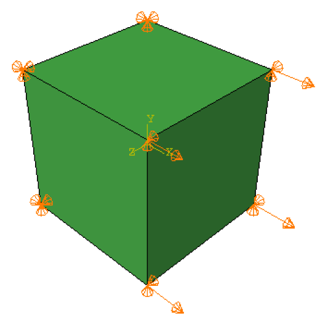}
		\caption{Single element (C3D8) model in uniaxial tension}
		\label{fig:fig5}
	\end{minipage}%
	\hspace{0.5cm}
	\begin{minipage}{0.4\linewidth}
		\includegraphics[width=\textwidth]{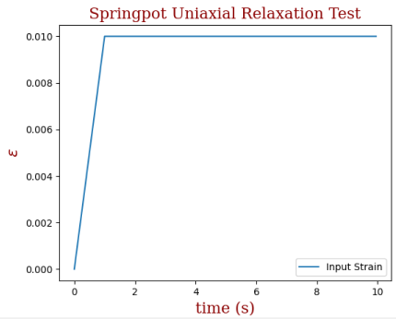}
		\caption{Relaxation test with maximum strain magnitude = 0.01}
		\label{fig:fig6}
	\end{minipage}%
	\hspace{0.5cm}
	\begin{minipage}{0.47\linewidth}
		\includegraphics[width=\textwidth]{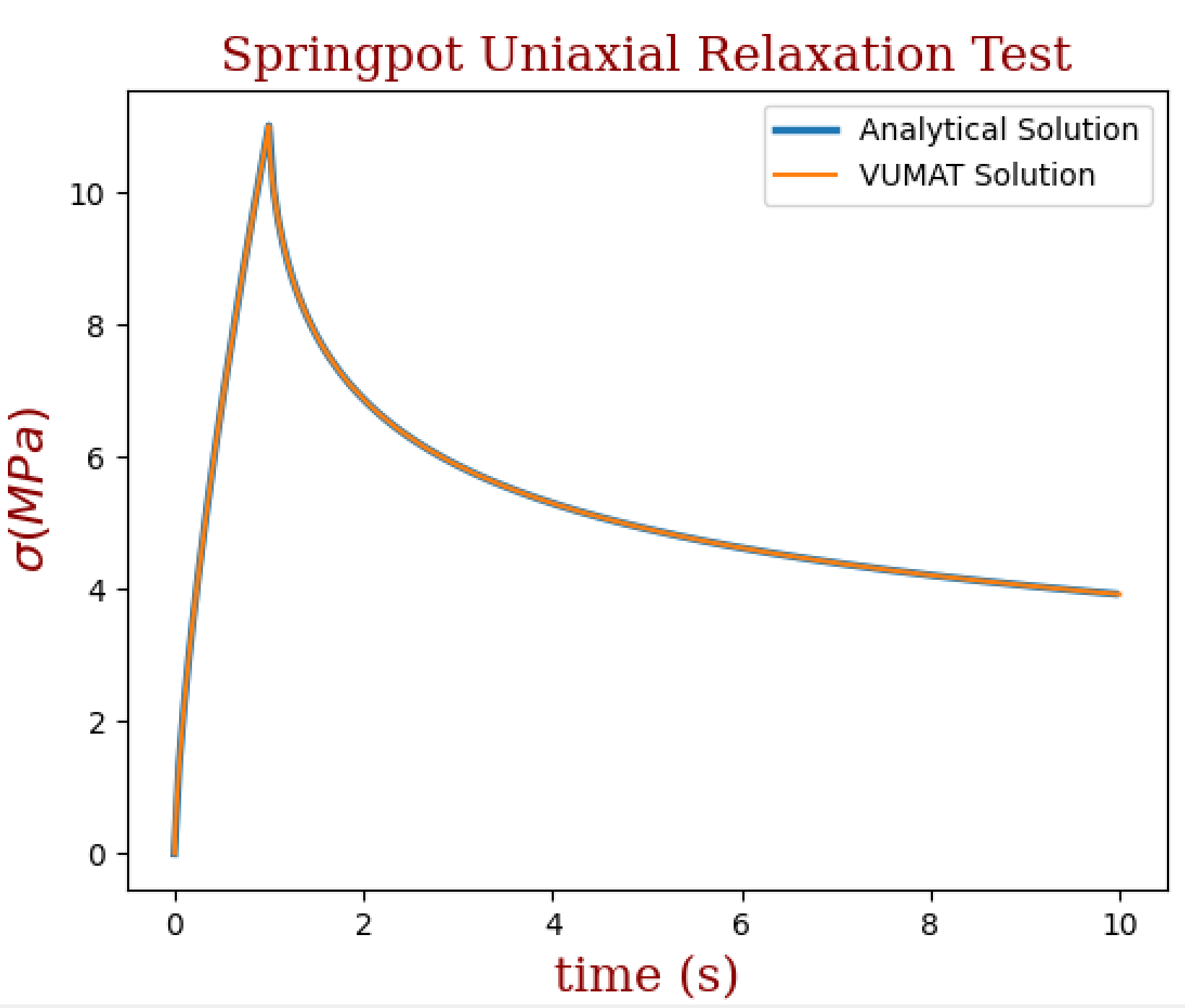}
		\caption{Comparison of the analytical solution vs VUMAT solution for the relaxation test (Figure 6.) of fractional viscoelastic model}
		\label{fig:compare}
	\end{minipage}
	\hspace{0.4cm}
	\begin{minipage}{0.47\linewidth}
	\includegraphics[width=\textwidth]{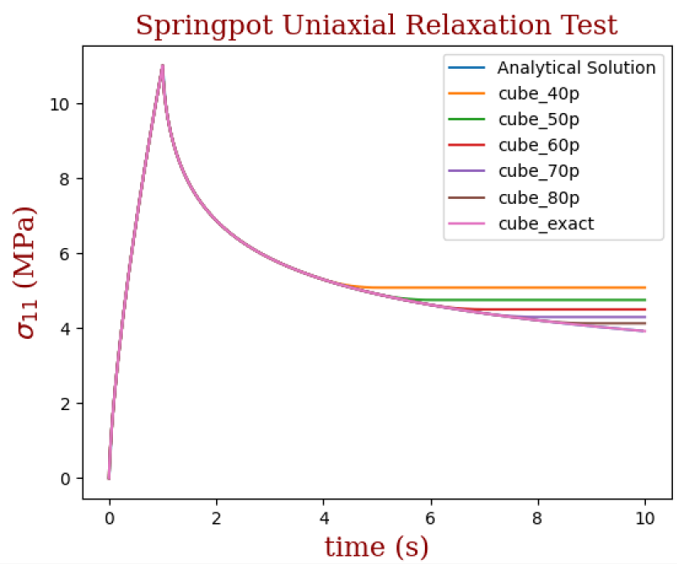}
	\caption{Reduced Memory solution for the single element uniaxial test case. The solution is computed for 40\% - 80\% of the total time. 40 \% memory length for a total time of 10 s implies $L = 0.4 * 10 = 4 s.$ }
	\label{fig:rm}
\end{minipage}
\end{figure}

\noindent The homogenized stress along the fiber direction for the RVE with 40\% volume fraction is shown in Figure 10. The stress along the fiber direction for different volume fractions is shown in Figure 11. The stress distribution for the six loading directions is shown in Figure 12. The simulations were performed with a memory length of 60 percent of the total simulation time. It can be seen that the RVE becomes stiffer with increasing volume fraction of axons. The output stress along the fiber direction, $\sigma_{11}$, for different design ids sampled by the optimization analysis for an RVE with 40 \% volume fraction is shown in Figure 13. A total of 800 iterations is performed during the optimization process achieving a minimum cost function value of $3 \times 10^{-5}$. The optimal fractional parameters are obtained similarly for each loading direction. Finally, an orthotropic fractional viscoelastic compliance matrix is formulated based on the homogenized RVE material properties as a function of the volume fraction of the RVE, see equation (19). An example of the orthotropic fractional properties is shown in Table 4. The Poisson's ratio in each direction is shown in Table 5. The parameters in Table 4 are cumbersome because of their units, $kPa \cdot ms^\beta$. Therefore, these parameters are transformed to the dimensions of shear modulus, $kPa$, using equation (18) [\onlinecite{sack2009impact}], where $\eta = 3.7 \hspace{1mm}kPa \cdot ms$ [\onlinecite{sack2009impact}]. The resulting \textsl{shear moduli} like parameters are shown in Table 6.

\begin{equation}
	E = (E_\beta * \eta^{-\beta})^{\frac{1}{1 - \beta}} \hspace{3mm}
\end{equation}

\begin{figure}[ht!]
	\begin{minipage}{0.3\linewidth}
		\includegraphics[width=\textwidth]{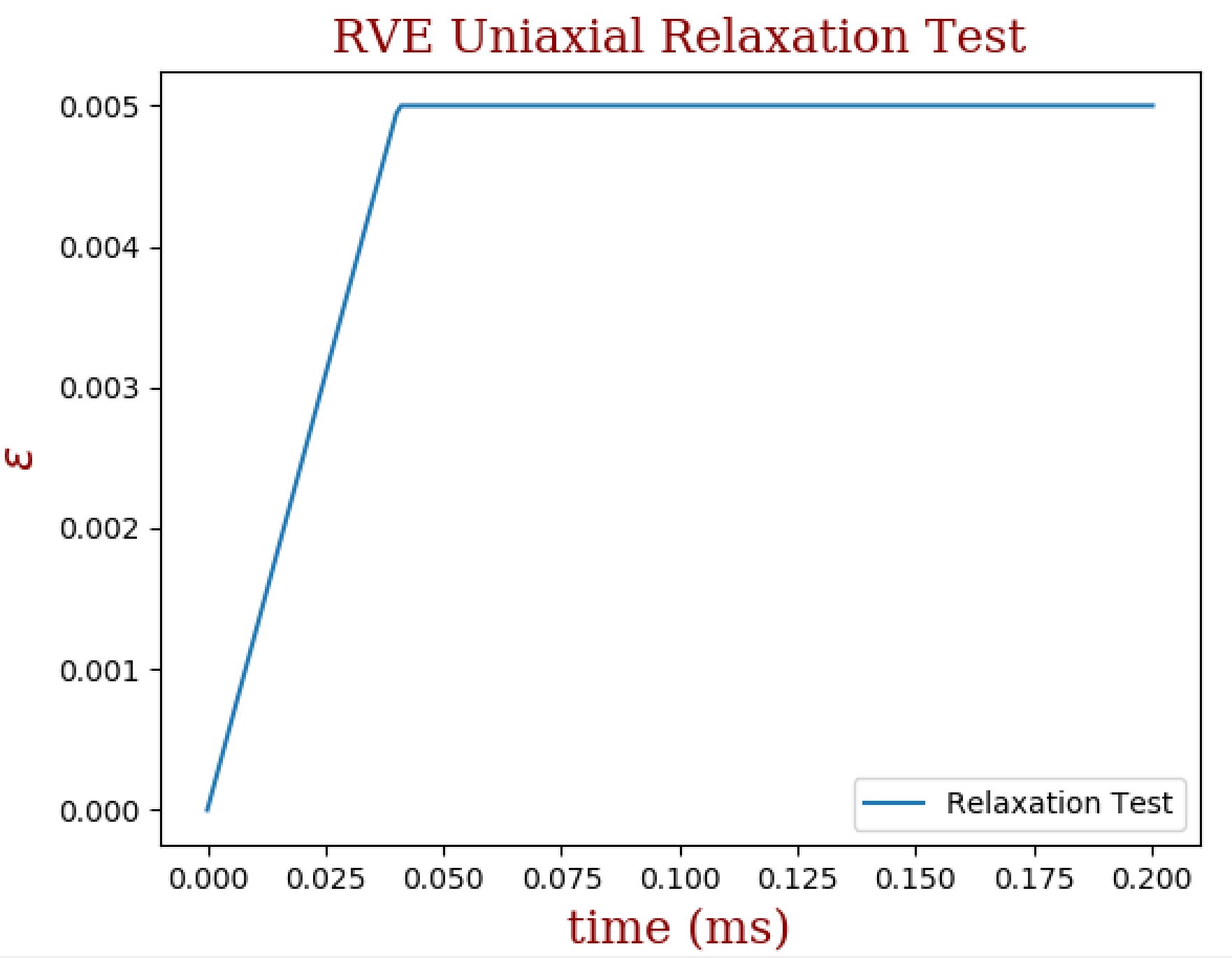}
		\caption{Relaxation test with maximum strain magnitude = 0.01. Total simulation time = $0.2025 \mu s$}
		\label{fig:figure1}
	\end{minipage}%
	\hspace{0.5cm}
	\begin{minipage}{0.30\linewidth}
		\includegraphics[width=\textwidth]{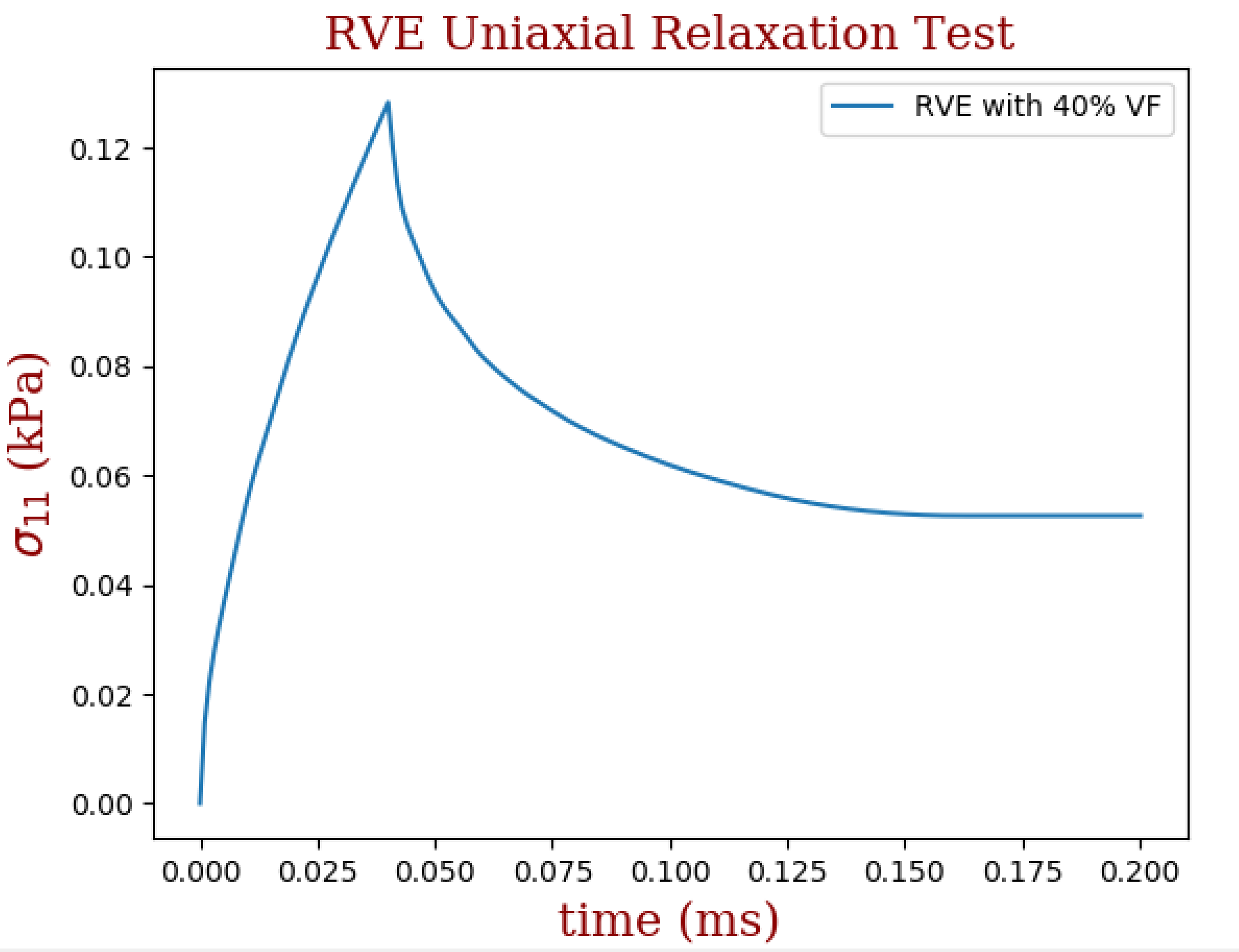}
		\caption{$\sigma_{11}$ along fiber direction for RVE 40\% axon volume fraction.}
		\label{fig:figure2}
	\end{minipage}%
	\hspace{0.5cm}
		\begin{minipage}{0.30\linewidth}
		\includegraphics[width=\textwidth]{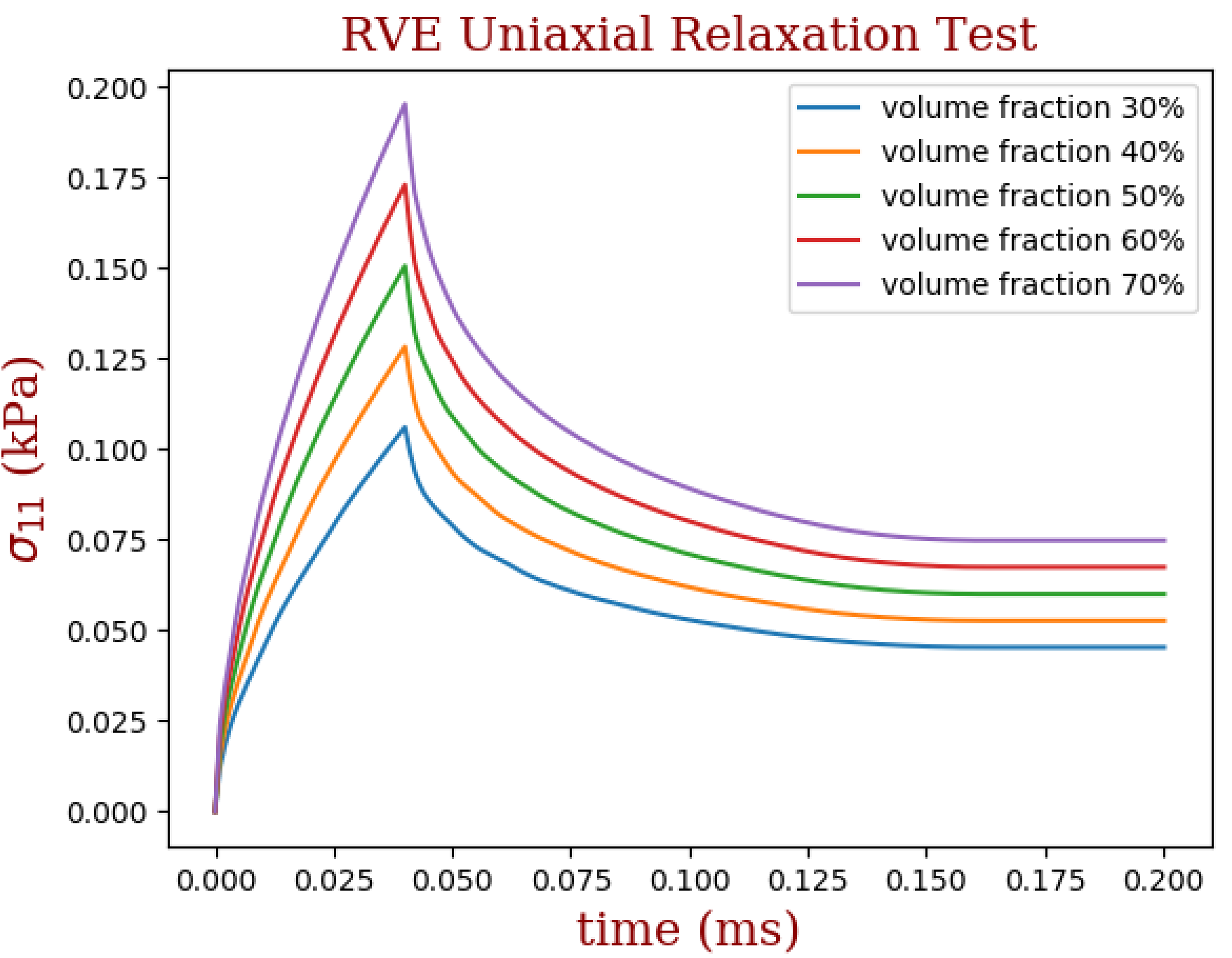}
		\caption{$\sigma_{11}$ along fiber direction for RVE with volume fraction ranging from 30\% - 70\%.}
		\label{fig:figure2}
	\end{minipage}%
	\hspace{0.5cm}
	\begin{minipage}{0.7\linewidth}
		\includegraphics[width=\textwidth]{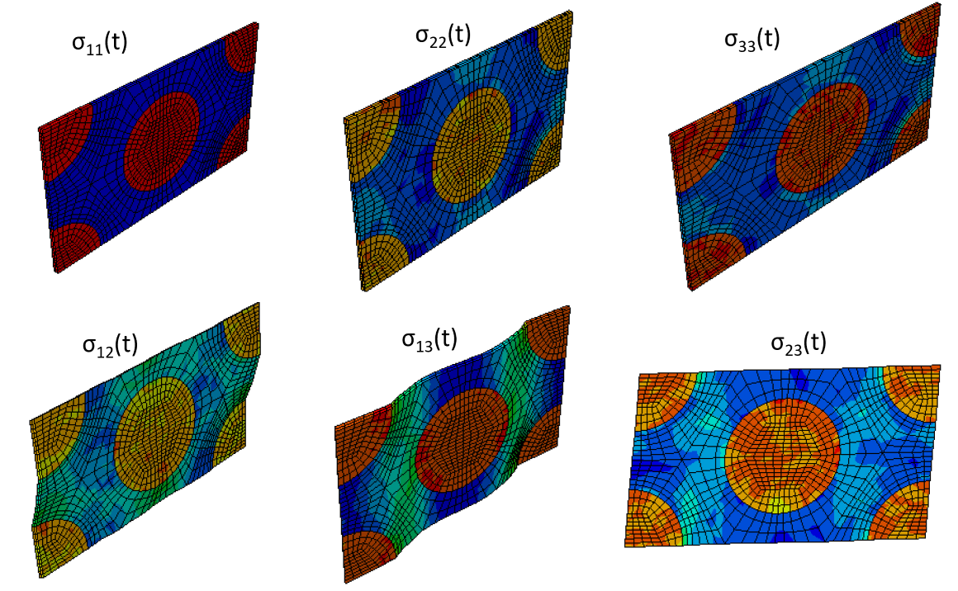}
		\caption{Stress and deformation plot for RVE with 40\% volume fraction in six loading directions.}
		\label{fig:figure3}
	\end{minipage}
\end{figure}

\begin{figure}[ht!]
	\includegraphics[scale=0.5]{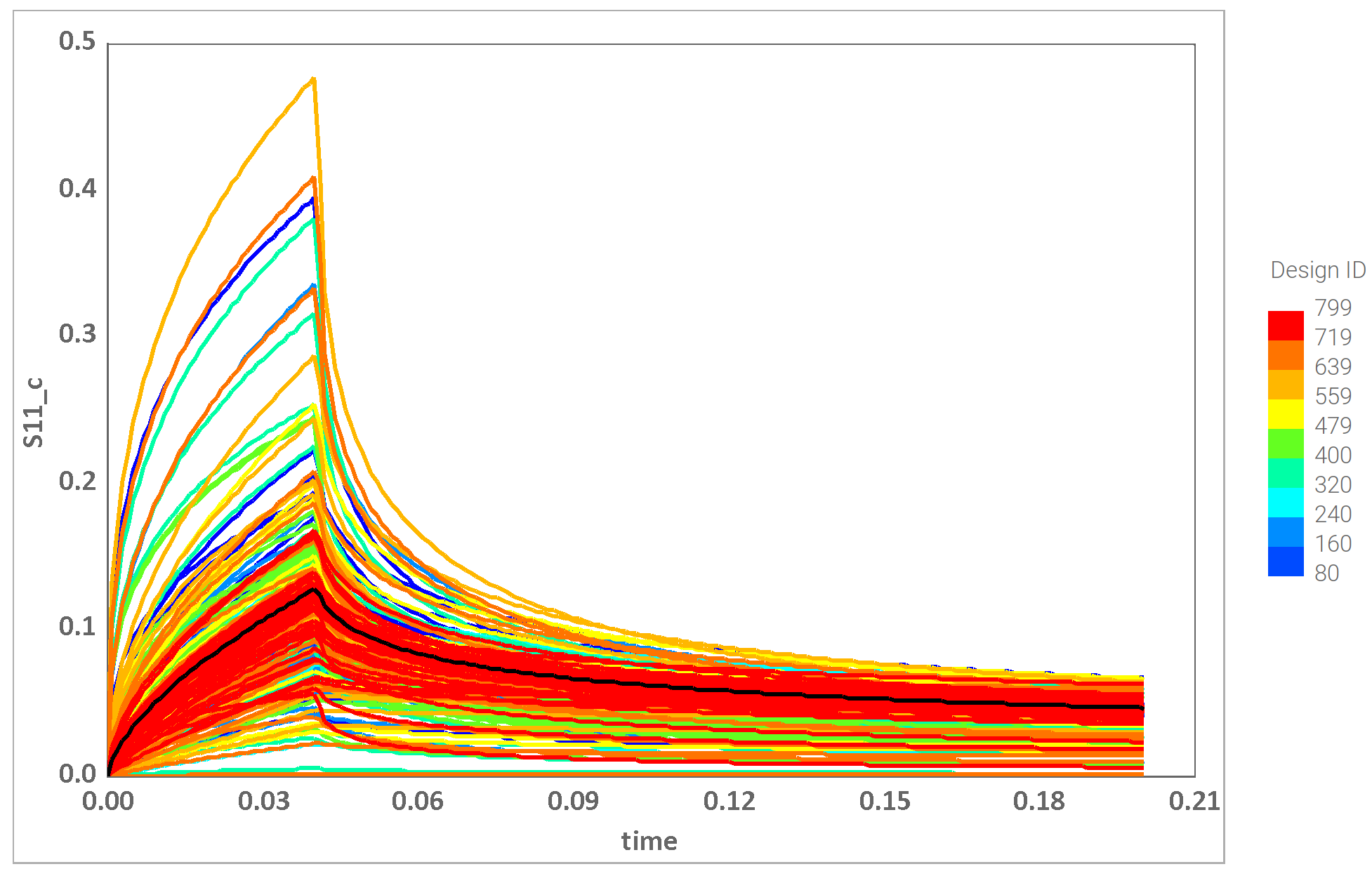}
	\caption{\label{opt}  Homogenized stress computed by the optimization algorithm for different design ids sampled by modeFRONTIER. The optimal curve with the lowest cost function is highlighted in black.}
\end{figure}

\def\doubleunderline#1{\underline{\underline{#1}}}
\begin{equation}
	\doubleunderline{S_\beta} = 
	\begin{bmatrix}
		\frac{1}{E_{11}}, \beta_{11} & -\frac{\nu_{12}}{E_{22}}, \beta_{22}  & -\frac{\nu_{12}}{E_{33}}, \beta_{33} & 0 & 0 & 0 \\ \\
		-\frac{\nu_{21}}{E_{11}}, \beta_{11} & \frac{1}{E_{22}}, \beta_{22} & -\frac{\nu_{23}}{E_{33}}, \beta_{33} & 0 & 0 & 0 \\ \\
		-\frac{\nu_{31}}{E_{11}}, \beta_{11} & -\frac{\nu_{32}}{E_{22}}, \beta_{22} & \frac{1}{E_{33}}, \beta_{33} & 0 & 0 & 0 \\ \\
		0 & 0 & 0 & \frac{1}{G_{23}}, \beta_{23} & 0 & 0 \\ \\
		0 & 0 & 0 & 0 & \frac{1}{G_{31}}, \beta_{31} & 0 \\ \\
		0 & 0 & 0 & 0 & 0 & \frac{1}{G_{12}}, \beta_{13}
	\end{bmatrix}
\end{equation}

\begin{table}[]
	\caption{\label{tab:table1}Homogenized fractional material properties for an RVE with a 40 \% axon volume fraction}
	\begin{ruledtabular}
	\begin{tabular}{cccccccc}
		Moduli $(kPa \cdot ms^{\beta})$ & Value    & $\beta$      & Value  & Moduli $(kPa \cdot ms^{\beta})$ & Value   & $\beta$      & Value  \\ \hline
		$E_{11}$                          & 7.0346  & $\beta_{11}$ & 0.3719 & $G_{12}$                          & 1.4595 & $\beta_{12}$ & 0.1912 \\ \hline
		$E_{22}$                          & 7.600 & $\beta_{22}$ & 0.180 & $G_{23}$                          & 1.442   & $\beta_{23}$ & 0.1792 \\ \hline
		$E_{33}$                          & 7.4392 & $\beta_{33}$ & 0.190 & $G_{13}$                          & 1.2790 & $\beta_{13}$ & 0.2548 \\
	\end{tabular}
	\end{ruledtabular}
\end{table}

\begin{table}[tbp]
	\begin{minipage}[t]{0.3\textwidth}
		\caption{Poisson's ratio for the orthotropic fiber composite RVE of axons in ECM}
		\label{tab:multiples2}
		\begin{ruledtabular}
			\begin{tabular}{cc}
		$\nu$    & value    \\ \hline
		$\nu_{12}$ & 0.2995 \\ \hline
		$\nu_{13}$ & 0.3060  \\ \hline
		$\nu_{21}$ & 0.2418   \\ \hline
		$\nu_{23}$ & 0.3156  \\ \hline
		$\nu_{31}$ & 0.2350  \\ \hline
		$\nu_{32}$ & 0.3000   \\ 
			\end{tabular}
		\end{ruledtabular}
	\end{minipage}\hspace{7mm}
	\begin{minipage}[t]{0.3\textwidth}
		\caption{Fractional material properties transformed to shear moduli like parameters [\onlinecite{sack2009impact}]}
		\label{tab:multiples3}
		\begin{ruledtabular}
			\begin{tabular}{ccc}
			 E (kPa) & value     \\ \hline
			$E_{11}$   & 10.291 \\ \hline
			$E_{22}$    & 8.900  \\ \hline
			$E_{33}$    & 8.763  \\ \hline
			$G_{23}$    & 1.171  \\ \hline
			$G_{31}$    & 0.889  \\ \hline
			$G_{12}$    & 1.174  \\ 
			\end{tabular}
		\end{ruledtabular}
	\end{minipage}
\end{table}

\section{Conclusion}

Effectively characterizing the mechanical response of brain white matter across multiple spatial and temporal scales is inherently difficult. While transforming frequency related shear moduli parameters into a Prony series fit is a practical solution, the Prony parameters are somewhat cumbersome and hard to interpret in terms of their physical significance. In this study, a fractional viscoelastic model of the axons and ECM is developed in the time domain. The fractional model is comparatively elegant and its parameters easier to interpret. Using an optimization scheme, homogenized material properties for the RVE are extracted for a relaxation displacement controlled boundary condition. The model can be extended to study different loading conditions within a viscoelastic domain and yield insight into the correlations between axon volume fractions and the different directional moduli. The long-term goal of this study is to build a database of material parameters for different load cases, axon volume fractions, and incorporate the properties into tissue-scale models which can be directly compared to experiments. 

\begin{acknowledgments}
	Support was provided by NSF Grants CMMI-1436743, CMMI-1437113, CMMI-1762774, CMMI-1763005, and the R.A. Pritzker endowed chair.
\end{acknowledgments}

\nocite{*}
\bibliography{aipsamp}

\providecommand{\noopsort}[1]{}\providecommand{\singleletter}[1]{#1}%
\begin{thebibliography}{23}%
\makeatletter
\providecommand \@ifxundefined [1]{%
 \@ifx{#1\undefined}
}%
\providecommand \@ifnum [1]{%
 \ifnum #1\expandafter \@firstoftwo
 \else \expandafter \@secondoftwo
 \fi
}%
\providecommand \@ifx [1]{%
 \ifx #1\expandafter \@firstoftwo
 \else \expandafter \@secondoftwo
 \fi
}%
\providecommand \natexlab [1]{#1}%
\providecommand \enquote  [1]{``#1''}%
\providecommand \bibnamefont  [1]{#1}%
\providecommand \bibfnamefont [1]{#1}%
\providecommand \citenamefont [1]{#1}%
\providecommand \href@noop [0]{\@secondoftwo}%
\providecommand \href [0]{\begingroup \@sanitize@url \@href}%
\providecommand \@href[1]{\@@startlink{#1}\@@href}%
\providecommand \@@href[1]{\endgroup#1\@@endlink}%
\providecommand \@sanitize@url [0]{\catcode `\\12\catcode `\$12\catcode
  `\&12\catcode `\#12\catcode `\^12\catcode `\_12\catcode `\%12\relax}%
\providecommand \@@startlink[1]{}%
\providecommand \@@endlink[0]{}%
\providecommand \url  [0]{\begingroup\@sanitize@url \@url }%
\providecommand \@url [1]{\endgroup\@href {#1}{\urlprefix }}%
\providecommand \urlprefix  [0]{URL }%
\providecommand \Eprint [0]{\href }%
\providecommand \doibase [0]{https://doi.org/}%
\providecommand \selectlanguage [0]{\@gobble}%
\providecommand \bibinfo  [0]{\@secondoftwo}%
\providecommand \bibfield  [0]{\@secondoftwo}%
\providecommand \translation [1]{[#1]}%
\providecommand \BibitemOpen [0]{}%
\providecommand \bibitemStop [0]{}%
\providecommand \bibitemNoStop [0]{.\EOS\space}%
\providecommand \EOS [0]{\spacefactor3000\relax}%
\providecommand \BibitemShut  [1]{\csname bibitem#1\endcsname}%
\let\auto@bib@innerbib\@empty
\bibitem [{\citenamefont {Aare}\ and\ \citenamefont
  {Kleiven}(2007)}]{AARE2007596}%
  \BibitemOpen
  \bibfield  {author} {\bibinfo {author} {\bibfnamefont {M.}~\bibnamefont
  {Aare}}\ and\ \bibinfo {author} {\bibfnamefont {S.}~\bibnamefont {Kleiven}},\
  }\bibfield  {title} {\enquote {\bibinfo {title} {Evaluation of head response
  to ballistic helmet impacts using the finite element method},}\ }\href
  {https://doi.org/https://doi.org/10.1016/j.ijimpeng.2005.08.001} {\bibfield
  {journal} {\bibinfo  {journal} {International Journal of Impact Engineering}\
  }\textbf {\bibinfo {volume} {34}},\ \bibinfo {pages} {596--608} (\bibinfo
  {year} {2007})}\BibitemShut {NoStop}%
\bibitem [{\citenamefont {Pan}\ \emph {et~al.}(2013)\citenamefont {Pan},
  \citenamefont {Sullivan}, \citenamefont {Shreiber},\ and\ \citenamefont
  {Pelegri}}]{pan2013finite}%
  \BibitemOpen
  \bibfield  {author} {\bibinfo {author} {\bibfnamefont {Y.}~\bibnamefont
  {Pan}}, \bibinfo {author} {\bibfnamefont {D.}~\bibnamefont {Sullivan}},
  \bibinfo {author} {\bibfnamefont {D.~I.}\ \bibnamefont {Shreiber}},\ and\
  \bibinfo {author} {\bibfnamefont {A.~A.}\ \bibnamefont {Pelegri}},\
  }\bibfield  {title} {\enquote {\bibinfo {title} {Finite element modeling of
  cns white matter kinematics: use of a 3d rve to determine material
  properties},}\ }\href@noop {} {\bibfield  {journal} {\bibinfo  {journal}
  {Frontiers in bioengineering and biotechnology}\ }\textbf {\bibinfo {volume}
  {1}},\ \bibinfo {pages} {19} (\bibinfo {year} {2013})}\BibitemShut {NoStop}%
\bibitem [{\citenamefont {Abolfathi}\ \emph {et~al.}(2009)\citenamefont
  {Abolfathi}, \citenamefont {Naik}, \citenamefont {Sotudeh~Chafi},
  \citenamefont {Karami},\ and\ \citenamefont {Ziejewski}}]{abolfathi2009}%
  \BibitemOpen
  \bibfield  {author} {\bibinfo {author} {\bibfnamefont {N.}~\bibnamefont
  {Abolfathi}}, \bibinfo {author} {\bibfnamefont {A.}~\bibnamefont {Naik}},
  \bibinfo {author} {\bibfnamefont {M.}~\bibnamefont {Sotudeh~Chafi}}, \bibinfo
  {author} {\bibfnamefont {G.}~\bibnamefont {Karami}},\ and\ \bibinfo {author}
  {\bibfnamefont {M.}~\bibnamefont {Ziejewski}},\ }\bibfield  {title} {\enquote
  {\bibinfo {title} {A micromechanical procedure for modelling the anisotropic
  mechanical properties of brain white matter},}\ }\href@noop {} {\bibfield
  {journal} {\bibinfo  {journal} {Computer Methods in Biomechanics and
  Biomedical Engineering}\ }\textbf {\bibinfo {volume} {12}},\ \bibinfo {pages}
  {3} (\bibinfo {year} {2009})}\BibitemShut {NoStop}%
\bibitem [{\citenamefont {Sullivan}\ \emph {et~al.}(2021)\citenamefont
  {Sullivan}, \citenamefont {Wu}, \citenamefont {Gallo}, \citenamefont
  {Naughton}, \citenamefont {Georgiadis},\ and\ \citenamefont
  {Pelegri}}]{sullivan2021}%
  \BibitemOpen
  \bibfield  {author} {\bibinfo {author} {\bibfnamefont {D.~J.}\ \bibnamefont
  {Sullivan}}, \bibinfo {author} {\bibfnamefont {X.}~\bibnamefont {Wu}},
  \bibinfo {author} {\bibfnamefont {N.~R.}\ \bibnamefont {Gallo}}, \bibinfo
  {author} {\bibfnamefont {N.~M.}\ \bibnamefont {Naughton}}, \bibinfo {author}
  {\bibfnamefont {J.~G.}\ \bibnamefont {Georgiadis}},\ and\ \bibinfo {author}
  {\bibfnamefont {A.~A.}\ \bibnamefont {Pelegri}},\ }\bibfield  {title}
  {\enquote {\bibinfo {title} {Sensitivity analysis of effective transverse
  shear viscoelastic and diffusional properties of myelinated white matter},}\
  }\href@noop {} {\bibfield  {journal} {\bibinfo  {journal} {Physics in
  Medicine and Biology}\ }\textbf {\bibinfo {volume} {66}},\ \bibinfo {pages}
  {3} (\bibinfo {year} {2021})}\BibitemShut {NoStop}%
\bibitem [{\citenamefont {Arbogast}\ and\ \citenamefont
  {Margulies}(1998)}]{arbogast1998material}%
  \BibitemOpen
  \bibfield  {author} {\bibinfo {author} {\bibfnamefont {K.~B.}\ \bibnamefont
  {Arbogast}}\ and\ \bibinfo {author} {\bibfnamefont {S.~S.}\ \bibnamefont
  {Margulies}},\ }\bibfield  {title} {\enquote {\bibinfo {title} {Material
  characterization of the brainstem from oscillatory shear tests},}\
  }\href@noop {} {\bibfield  {journal} {\bibinfo  {journal} {Journal of
  biomechanics}\ }\textbf {\bibinfo {volume} {31}},\ \bibinfo {pages}
  {801--807} (\bibinfo {year} {1998})}\BibitemShut {NoStop}%
\bibitem [{\citenamefont {Meaney}(2003)}]{meaney2003}%
  \BibitemOpen
  \bibfield  {author} {\bibinfo {author} {\bibfnamefont {D.~F.}\ \bibnamefont
  {Meaney}},\ }\bibfield  {title} {\enquote {\bibinfo {title} {Relationship
  between structural modeling and hyperelastic material behavior: application
  to cns white matter},}\ }\href@noop {} {\bibfield  {journal} {\bibinfo
  {journal} {Biomechanics and modeling in mechanobiology}\ }\textbf {\bibinfo
  {volume} {1}},\ \bibinfo {pages} {279--293} (\bibinfo {year}
  {2003})}\BibitemShut {NoStop}%
\bibitem [{\citenamefont {Montanino}\ and\ \citenamefont
  {Kleiven}(2018)}]{montanino2018}%
  \BibitemOpen
  \bibfield  {author} {\bibinfo {author} {\bibfnamefont {A.}~\bibnamefont
  {Montanino}}\ and\ \bibinfo {author} {\bibfnamefont {S.}~\bibnamefont
  {Kleiven}},\ }\bibfield  {title} {\enquote {\bibinfo {title} {Utilizing a
  structural mechanics approach to assess the primary effects of injury loads
  onto the axon and its components},}\ }\href@noop {} {\bibfield  {journal}
  {\bibinfo  {journal} {Frontiers in neurology}\ ,\ \bibinfo {pages} {643}}
  (\bibinfo {year} {2018})}\BibitemShut {NoStop}%
\bibitem [{\citenamefont {Javid}, \citenamefont {Rezaei},\ and\ \citenamefont
  {Karami}(2014)}]{JAVID2014290}%
  \BibitemOpen
  \bibfield  {author} {\bibinfo {author} {\bibfnamefont {S.}~\bibnamefont
  {Javid}}, \bibinfo {author} {\bibfnamefont {A.}~\bibnamefont {Rezaei}},\ and\
  \bibinfo {author} {\bibfnamefont {G.}~\bibnamefont {Karami}},\ }\bibfield
  {title} {\enquote {\bibinfo {title} {A micromechanical procedure for
  viscoelastic characterization of the axons and ecm of the brainstem},}\
  }\href {https://doi.org/https://doi.org/10.1016/j.jmbbm.2013.11.010}
  {\bibfield  {journal} {\bibinfo  {journal} {Journal of the Mechanical
  Behavior of Biomedical Materials}\ }\textbf {\bibinfo {volume} {30}},\
  \bibinfo {pages} {290--299} (\bibinfo {year} {2014})}\BibitemShut {NoStop}%
\bibitem [{\citenamefont {Pan}, \citenamefont {Shreiber},\ and\ \citenamefont
  {Pelegri}(2011)}]{pan2011transition}%
  \BibitemOpen
  \bibfield  {author} {\bibinfo {author} {\bibfnamefont {Y.}~\bibnamefont
  {Pan}}, \bibinfo {author} {\bibfnamefont {D.~I.}\ \bibnamefont {Shreiber}},\
  and\ \bibinfo {author} {\bibfnamefont {A.~A.}\ \bibnamefont {Pelegri}},\
  }\bibfield  {title} {\enquote {\bibinfo {title} {A transition model for
  finite element simulation of kinematics of central nervous system white
  matter},}\ }\href@noop {} {\bibfield  {journal} {\bibinfo  {journal} {IEEE
  transactions on biomedical engineering}\ }\textbf {\bibinfo {volume} {58}},\
  \bibinfo {pages} {3443--3446} (\bibinfo {year} {2011})}\BibitemShut {NoStop}%
\bibitem [{\citenamefont {Sack}\ \emph {et~al.}(2009)\citenamefont {Sack},
  \citenamefont {Beierbach}, \citenamefont {Wuerfel}, \citenamefont {Klatt},
  \citenamefont {Hamhaber}, \citenamefont {Papazoglou}, \citenamefont
  {Martus},\ and\ \citenamefont {Braun}}]{sack2009impact}%
  \BibitemOpen
  \bibfield  {author} {\bibinfo {author} {\bibfnamefont {I.}~\bibnamefont
  {Sack}}, \bibinfo {author} {\bibfnamefont {B.}~\bibnamefont {Beierbach}},
  \bibinfo {author} {\bibfnamefont {J.}~\bibnamefont {Wuerfel}}, \bibinfo
  {author} {\bibfnamefont {D.}~\bibnamefont {Klatt}}, \bibinfo {author}
  {\bibfnamefont {U.}~\bibnamefont {Hamhaber}}, \bibinfo {author}
  {\bibfnamefont {S.}~\bibnamefont {Papazoglou}}, \bibinfo {author}
  {\bibfnamefont {P.}~\bibnamefont {Martus}},\ and\ \bibinfo {author}
  {\bibfnamefont {J.}~\bibnamefont {Braun}},\ }\bibfield  {title} {\enquote
  {\bibinfo {title} {The impact of aging and gender on brain
  viscoelasticity},}\ }\href@noop {} {\bibfield  {journal} {\bibinfo  {journal}
  {Neuroimage}\ }\textbf {\bibinfo {volume} {46}},\ \bibinfo {pages} {652--657}
  (\bibinfo {year} {2009})}\BibitemShut {NoStop}%
\bibitem [{\citenamefont {Sack}\ \emph {et~al.}(2013)\citenamefont {Sack},
  \citenamefont {J{\"o}hrens}, \citenamefont {W{\"u}rfel},\ and\ \citenamefont
  {Braun}}]{sack2013structure}%
  \BibitemOpen
  \bibfield  {author} {\bibinfo {author} {\bibfnamefont {I.}~\bibnamefont
  {Sack}}, \bibinfo {author} {\bibfnamefont {K.}~\bibnamefont {J{\"o}hrens}},
  \bibinfo {author} {\bibfnamefont {J.}~\bibnamefont {W{\"u}rfel}},\ and\
  \bibinfo {author} {\bibfnamefont {J.}~\bibnamefont {Braun}},\ }\bibfield
  {title} {\enquote {\bibinfo {title} {Structure-sensitive elastography: on the
  viscoelastic powerlaw behavior of in vivo human tissue in health and
  disease},}\ }\href@noop {} {\bibfield  {journal} {\bibinfo  {journal} {Soft
  matter}\ }\textbf {\bibinfo {volume} {9}},\ \bibinfo {pages} {5672--5680}
  (\bibinfo {year} {2013})}\BibitemShut {NoStop}%
\bibitem [{\citenamefont {Nicolas}\ \emph {et~al.}(2018)\citenamefont
  {Nicolas}, \citenamefont {Calle}, \citenamefont {Nicolle}, \citenamefont
  {Mitton},\ and\ \citenamefont {Remenieras}}]{nicolas2018biomechanical}%
  \BibitemOpen
  \bibfield  {author} {\bibinfo {author} {\bibfnamefont {E.}~\bibnamefont
  {Nicolas}}, \bibinfo {author} {\bibfnamefont {S.}~\bibnamefont {Calle}},
  \bibinfo {author} {\bibfnamefont {S.}~\bibnamefont {Nicolle}}, \bibinfo
  {author} {\bibfnamefont {D.}~\bibnamefont {Mitton}},\ and\ \bibinfo {author}
  {\bibfnamefont {J.-P.}\ \bibnamefont {Remenieras}},\ }\bibfield  {title}
  {\enquote {\bibinfo {title} {Biomechanical characterization of ex vivo human
  brain using ultrasound shear wave spectroscopy},}\ }\href@noop {} {\bibfield
  {journal} {\bibinfo  {journal} {Ultrasonics}\ }\textbf {\bibinfo {volume}
  {84}},\ \bibinfo {pages} {119--125} (\bibinfo {year} {2018})}\BibitemShut
  {NoStop}%
\bibitem [{\citenamefont {Kurt}\ \emph {et~al.}(2019)\citenamefont {Kurt},
  \citenamefont {Wu}, \citenamefont {Laksari}, \citenamefont {Ozkaya},
  \citenamefont {Suar}, \citenamefont {Lv}, \citenamefont {Epperson},
  \citenamefont {Epperson}, \citenamefont {Sawyer}, \citenamefont {Camarillo}
  \emph {et~al.}}]{kurt2019optimization}%
  \BibitemOpen
  \bibfield  {author} {\bibinfo {author} {\bibfnamefont {M.}~\bibnamefont
  {Kurt}}, \bibinfo {author} {\bibfnamefont {L.}~\bibnamefont {Wu}}, \bibinfo
  {author} {\bibfnamefont {K.}~\bibnamefont {Laksari}}, \bibinfo {author}
  {\bibfnamefont {E.}~\bibnamefont {Ozkaya}}, \bibinfo {author} {\bibfnamefont
  {Z.~M.}\ \bibnamefont {Suar}}, \bibinfo {author} {\bibfnamefont
  {H.}~\bibnamefont {Lv}}, \bibinfo {author} {\bibfnamefont {K.}~\bibnamefont
  {Epperson}}, \bibinfo {author} {\bibfnamefont {K.}~\bibnamefont {Epperson}},
  \bibinfo {author} {\bibfnamefont {A.~M.}\ \bibnamefont {Sawyer}}, \bibinfo
  {author} {\bibfnamefont {D.}~\bibnamefont {Camarillo}}, \emph {et~al.},\
  }\bibfield  {title} {\enquote {\bibinfo {title} {Optimization of a
  multifrequency magnetic resonance elastography protocol for the human
  brain},}\ }\href@noop {} {\bibfield  {journal} {\bibinfo  {journal} {Journal
  of Neuroimaging}\ }\textbf {\bibinfo {volume} {29}},\ \bibinfo {pages}
  {440--446} (\bibinfo {year} {2019})}\BibitemShut {NoStop}%
\bibitem [{\citenamefont {Bonfanti}\ \emph {et~al.}(2020)\citenamefont
  {Bonfanti}, \citenamefont {Kaplan}, \citenamefont {Charras},\ and\
  \citenamefont {Kabla}}]{bonfanti2020fractional}%
  \BibitemOpen
  \bibfield  {author} {\bibinfo {author} {\bibfnamefont {A.}~\bibnamefont
  {Bonfanti}}, \bibinfo {author} {\bibfnamefont {J.~L.}\ \bibnamefont
  {Kaplan}}, \bibinfo {author} {\bibfnamefont {G.}~\bibnamefont {Charras}},\
  and\ \bibinfo {author} {\bibfnamefont {A.}~\bibnamefont {Kabla}},\ }\bibfield
   {title} {\enquote {\bibinfo {title} {Fractional viscoelastic models for
  power-law materials},}\ }\href@noop {} {\bibfield  {journal} {\bibinfo
  {journal} {Soft Matter}\ }\textbf {\bibinfo {volume} {16}},\ \bibinfo {pages}
  {6002--6020} (\bibinfo {year} {2020})}\BibitemShut {NoStop}%
\bibitem [{\citenamefont {Alotta}\ \emph {et~al.}(2017)\citenamefont {Alotta},
  \citenamefont {Barrera}, \citenamefont {Cocks},\ and\ \citenamefont
  {Paola}}]{alotta2017behavior}%
  \BibitemOpen
  \bibfield  {author} {\bibinfo {author} {\bibfnamefont {G.}~\bibnamefont
  {Alotta}}, \bibinfo {author} {\bibfnamefont {O.}~\bibnamefont {Barrera}},
  \bibinfo {author} {\bibfnamefont {A.~C.}\ \bibnamefont {Cocks}},\ and\
  \bibinfo {author} {\bibfnamefont {M.~D.}\ \bibnamefont {Paola}},\ }\bibfield
  {title} {\enquote {\bibinfo {title} {On the behavior of a three-dimensional
  fractional viscoelastic constitutive model},}\ }\href@noop {} {\bibfield
  {journal} {\bibinfo  {journal} {Meccanica}\ }\textbf {\bibinfo {volume}
  {52}},\ \bibinfo {pages} {2127--2142} (\bibinfo {year} {2017})}\BibitemShut
  {NoStop}%
\bibitem [{\citenamefont {Alotta}\ \emph {et~al.}(2018)\citenamefont {Alotta},
  \citenamefont {Barrera}, \citenamefont {Cocks},\ and\ \citenamefont
  {Di~Paola}}]{alotta2018finite}%
  \BibitemOpen
  \bibfield  {author} {\bibinfo {author} {\bibfnamefont {G.}~\bibnamefont
  {Alotta}}, \bibinfo {author} {\bibfnamefont {O.}~\bibnamefont {Barrera}},
  \bibinfo {author} {\bibfnamefont {A.}~\bibnamefont {Cocks}},\ and\ \bibinfo
  {author} {\bibfnamefont {M.}~\bibnamefont {Di~Paola}},\ }\bibfield  {title}
  {\enquote {\bibinfo {title} {The finite element implementation of 3d
  fractional viscoelastic constitutive models},}\ }\href@noop {} {\bibfield
  {journal} {\bibinfo  {journal} {Finite Elements in Analysis and Design}\
  }\textbf {\bibinfo {volume} {146}},\ \bibinfo {pages} {28--41} (\bibinfo
  {year} {2018})}\BibitemShut {NoStop}%
\bibitem [{199(1999{\natexlab{a}})}]{199941}%
  \BibitemOpen
  \bibfield  {title} {\enquote {\bibinfo {title} {Chapter 2 - fractional
  derivatives and integrals},}\ }in\ \href
  {https://doi.org/https://doi.org/10.1016/S0076-5392(99)80021-6} {\emph
  {\bibinfo {booktitle} {Fractional Differential Equations}}},\ \bibinfo
  {series} {Mathematics in Science and Engineering}, Vol.\ \bibinfo {volume}
  {198},\ \bibinfo {editor} {edited by\ \bibinfo {editor} {\bibfnamefont
  {I.}~\bibnamefont {Podlubny}}}\ (\bibinfo  {publisher} {Elsevier},\ \bibinfo
  {year} {1999})\ pp.\ \bibinfo {pages} {41--119}\BibitemShut {NoStop}%
\bibitem [{199(1999{\natexlab{b}})}]{1999199}%
  \BibitemOpen
  \bibfield  {title} {\enquote {\bibinfo {title} {Chapter 7 - numerical
  evaluation of fractional derivatives},}\ }in\ \href
  {https://doi.org/https://doi.org/10.1016/S0076-5392(99)80026-5} {\emph
  {\bibinfo {booktitle} {Fractional Differential Equations}}},\ \bibinfo
  {series} {Mathematics in Science and Engineering}, Vol.\ \bibinfo {volume}
  {198},\ \bibinfo {editor} {edited by\ \bibinfo {editor} {\bibfnamefont
  {I.}~\bibnamefont {Podlubny}}}\ (\bibinfo  {publisher} {Elsevier},\ \bibinfo
  {year} {1999})\ pp.\ \bibinfo {pages} {199--221}\BibitemShut {NoStop}%
\bibitem [{\citenamefont {Tian}\ \emph {et~al.}(2019)\citenamefont {Tian},
  \citenamefont {Qi}, \citenamefont {Chao}, \citenamefont {Liang},\ and\
  \citenamefont {Fu}}]{tian2019periodic}%
  \BibitemOpen
  \bibfield  {author} {\bibinfo {author} {\bibfnamefont {W.}~\bibnamefont
  {Tian}}, \bibinfo {author} {\bibfnamefont {L.}~\bibnamefont {Qi}}, \bibinfo
  {author} {\bibfnamefont {X.}~\bibnamefont {Chao}}, \bibinfo {author}
  {\bibfnamefont {J.}~\bibnamefont {Liang}},\ and\ \bibinfo {author}
  {\bibfnamefont {M.}~\bibnamefont {Fu}},\ }\bibfield  {title} {\enquote
  {\bibinfo {title} {Periodic boundary condition and its numerical
  implementation algorithm for the evaluation of effective mechanical
  properties of the composites with complicated micro-structures},}\
  }\href@noop {} {\bibfield  {journal} {\bibinfo  {journal} {Composites Part B:
  Engineering}\ }\textbf {\bibinfo {volume} {162}},\ \bibinfo {pages} {1--10}
  (\bibinfo {year} {2019})}\BibitemShut {NoStop}%
\bibitem [{\citenamefont {Omairey}, \citenamefont {Dunning},\ and\
  \citenamefont {Sriramula}(2019)}]{omairey2019development}%
  \BibitemOpen
  \bibfield  {author} {\bibinfo {author} {\bibfnamefont {S.~L.}\ \bibnamefont
  {Omairey}}, \bibinfo {author} {\bibfnamefont {P.~D.}\ \bibnamefont
  {Dunning}},\ and\ \bibinfo {author} {\bibfnamefont {S.}~\bibnamefont
  {Sriramula}},\ }\bibfield  {title} {\enquote {\bibinfo {title} {Development
  of an abaqus plugin tool for periodic rve homogenisation},}\ }\href@noop {}
  {\bibfield  {journal} {\bibinfo  {journal} {Engineering with Computers}\
  }\textbf {\bibinfo {volume} {35}},\ \bibinfo {pages} {567--577} (\bibinfo
  {year} {2019})}\BibitemShut {NoStop}%
\bibitem [{\citenamefont {Hibbitt}\ and\ \citenamefont {Inc}(2012)}]{abq}%
  \BibitemOpen
  \bibfield  {author} {\bibinfo {author} {\bibfnamefont {K.}~\bibnamefont
  {Hibbitt}}\ and\ \bibinfo {author} {\bibfnamefont {S.}~\bibnamefont {Inc}},\
  }\href@noop {} {\bibfield  {journal} {\bibinfo  {journal} {ABAQUS, ABAQUS
  Theory Manual}\ }\textbf {\bibinfo {volume} {94}},\ \bibinfo {pages} {262}
  (\bibinfo {year} {2012})}\BibitemShut {NoStop}%
\bibitem [{\citenamefont {ESTECO}()}]{mf}%
  \BibitemOpen
  \bibfield  {author} {\bibinfo {author} {\bibnamefont {ESTECO}},\ }\bibfield
  {title} {\enquote {\bibinfo {title} {{modeFRONTIER} - multi objective
  optimization design environment, www.esteco.com},}\ }\href@noop {} {\
  }\BibitemShut {NoStop}%
\bibitem [{\citenamefont {Hesammokri}(2019)}]{hesammokri2019implementation}%
  \BibitemOpen
  \bibfield  {author} {\bibinfo {author} {\bibfnamefont {P.}~\bibnamefont
  {Hesammokri}},\ }\emph {\bibinfo {title} {Implementation of fractional order
  viscoelastic models to finite element method}},\ \href@noop {} {Master's
  thesis},\ \bibinfo  {school} {Middle East Technical University} (\bibinfo
  {year} {2019})\BibitemShut {NoStop}%
\end{thebibliography}%

\end{document}